# Generative Adversarial Nets for Information Retrieval: Fundamentals and Advances

Weinan Zhang
Shanghai Jiao Tong University
wnzhang@sjtu.edu.cn

## ABSTRACT

Generative adversarial nets (GANs) have been widely studied during the recent development of deep learning and unsupervised learning. With an adversarial training mechanism, GAN manages to train a generative model to fit the underlying unknown real data distribution under the guidance of the discriminative model estimating whether a data instance is real or generated. Such a framework is originally proposed for fitting continuous data distribution such as images, thus it is not straightforward to be directly applied to information retrieval scenarios where the data is mostly discrete, such as IDs, text and graphs. In this tutorial, we focus on discussing the GAN techniques and the variants on discrete data fitting in various information retrieval scenarios. (i) We introduce the fundamentals of GAN framework and its theoretic properties; (ii) we carefully study the promising solutions to extend GAN onto discrete data generation; (iii) we introduce IRGAN, the fundamental GAN framework of fitting single ID data distribution and the direct application on information retrieval; (iv) we further discuss the task of sequential discrete data generation tasks, e.g., text generation, and the corresponding GAN solutions; (v) we present the most recent work on graph/network data fitting with node embedding techniques by GANs. Meanwhile, we also introduce the relevant open-source platforms such as IRGAN and Texygen to help audience conduct research experiments on GANs in information retrieval. Finally, we conclude this tutorial with a comprehensive summarization and a prospect of further research directions for GANs in information retrieval.

## 1 BACKGROUND AND MOTIVATIONS

Information retrieval (IR) refers to the task of returning a list of information items (e.g., documents, products or answers) given a particular information need (e.g., represented by query keywords, user profile or question sentence) [24]. For the methodologies of IR, in general there are two schools of thinking, namely (i) the generative modeling methods, which assume a generative process from the query to the document (or the inverse one) and build generative models to maximize the observation likelihood [20, 28], and (ii) the discriminative modeling methods, which directly build model to score each candidate pair of information need and item and then train the model in a regression or ranking framework [3, 16]. For both directions, machine learning techniques based on big data have been playing an important role.

On the other hand, the recent development and success of deep learning [14] have made it applied to various fields including information retrieval (IR). In fact, during the recent three years deep learning has to-some-extent been revolutionizing the IR area, involving the IR applications of web search [9, 21], recommender systems [1, 6], sentiment analysis [25] and question answering [19] etc. The Neural Information Retrieval (NeuIR) workshop[1] has become one of the most popular workshops in SIGIR 2016 and 2017.

However, most above deep learning work is merely focused on replacing the IR scoring functions with some carefully designed neural network architectures, i.e., the majority of the work on deep learning for IR is on discriminative models [9, 11, 21]. Therefore, one would think whether deep generative models could benefit IR solutions and, furthermore, whether the merge of generative and discriminative models could help each other towards a better final solution.

Generative adversarial nets (GANs) [8] are a novel deep unsupervised learning framework which has attracted much attention since its birth in 2014 [7]. In GAN framework, a generative model tries to map a randomly sampled noise vector to a data instance (e.g., an image) while a discriminative model is trained to judge whether a data instance is real or generated. With the interplay between the generative and discriminative models, there theoretically exists a Nash equilibrium where the generative model perfectly fits the underlying real data distribution while the discriminative model is totally fooled by the generated data.

With the inspiration from GANs, there has been some preliminary work trying applying GANs to solve information retrieval tasks since 2017 [24]. Regarding the user's preference distribution over the candidate documents as the underlying data distribution to fit, it is straightforward to build a GAN framework for IR tasks. The generative retrieval model $p(d|q,r)$ captures the conditional probability of the user (represented by her query $q$) selecting a candidate document $d$ given the relevance criteria $r$. The discriminative retrieval model $p(r|q,d)$ estimates whether the user likes the candidate document $d$. Such an interplay between the generative and discriminative retrieval models potentially unifies the two schools of thinking in IR [24].

---
[1]https://neu-ir.weebly.com/

From the data perspective, the major data format in IR is discrete tokens, e.g., IDs, text, or linked nodes in a network. Directly applying the original version of GANs onto such discrete data is infeasible because one cannot direct take gradient over the discrete data instance itself [8]. Thus, several branches of solutions have been proposed, including end-to-end Gumbel-softmax [12] and policy gradient based reinforcement learning [22].

In this tutorial, we introduce the very recent development of GAN techniques on information retrieval. We start from introducing the original GAN framework designed for continuous data generation and show its failure to be directly applied to discrete data generation. Then we introduce the several methodologies to bypassing the problem of taking gradient over discrete data and derive the corresponding models. Based on these fundamental techniques, we discuss the GAN models working on three different discrete data format in IR tasks, namely

- **IDs (tokens)**, for the most basic IR task and collaborative filtering for recommender systems;
- **Text (sequences of tokens)**, for sentence, poem, article generation and question answering;
- **Networks (graphs of tokens)**, for node embedding, node classification, link prediction and community detection etc;

where the fundamental thinking and different solutions will be discussed. Finally, we conclude this tutorial by summarizing the existing work on GANs for IR and discuss some promising future directions, e.g., the recently proposed Cooperative Training (CoT) [17].

## 2 OBJECTIVES

As a merge of information retrieval and machine learning, the target audience of this tutorial is sufficiently broad. And the main objective of this tutorial is to encourage the technical communications between informmation retrieval and machine learning, between application-oriented and methodology-oriented research, and between engineering and theory.

- For the researchers, engineers and students working on information retrieval such as text mining, natural language processing, graph pattern mining, network analysis, recommender systems etc., this tutorial offers a series of new methodologies which are potential to improve the model performance in their studied scenarios.
- For machine learning or artificial intelligence researchers, this tutorial helps them broad the scope by providing various application scenarios for generative adversarial nets and reinforcement learning.

With this tutorial, we hope the community would develop new research problems, novel methodologies and even new solution paradigms on information retrieval.

## 3 FORMAT AND DETAILED SCHEDULE

### 3.1 Tutorial Format

The tutorial is expected to be presented as a 3-hour lecture with slides and hangouts to the audience. It is free for the audience to raise questions during the lecture and there will be a half-hour QA session at the end of the tutorial.

### 3.2 Detailed Schedule

The planned 3-hour schedule of the tutorial is listed as follows.

**Introduction to GANs (35 mins).** We introduce the background and fundamentals of the original version of generative adversarial nets which are designed for continuous data generation. The theoretic Nash equilibrium of GAN will be discussed to help the audience understand the rationale behind such a minimax game. Also the recent advanced applications of GANs on image and speech generation scenarios will be presented to help the audience know about the wide applications of GANs. Finally, we reveal the problem of applying the original version of GANs to discrete data in information retrieval tasks, which lies in taking gradient over discrete tokens. Based on such a problem, we introduce several promising solutions including Gumbel-softmax and reinforcement learning.

**Reinforcement Learning (25 mins).** In order to extend GAN framework onto discrete data generation, we perform a brief review of reinforcement learning (RL). The basic idea of Markovian decision processes (MDPs), value function, policy gradient, REINFORCE algorithm [26], likelihood ratio trick and Monte Carlo search will be presented. A simple example will be provided to help the audience get familiar with the presented RL techniques.

**GANs for IR Tasks (35 mins).** With the previous fundamental delivered to the audience, we start to introduce the basic GAN models to the simplest IR tasks, where the information items are just represented by IDs (e.g., document ID or product ID without detailed information like text or multi-field features) and each query-document pair has been represented by a set of categorical and continuous features. Such a simple setting is illustrated in Figure 1. The IRGAN framework [24] is the central content of this part. We will also present more details beyond the IRGAN paper itself, e.g., the reward function design and importance sampling tricks. Furthermore, we will come to the ad hoc text retrieval task and link the GAN mechanism to (pseudo) relevance feedback and present the most recent work on this direction.

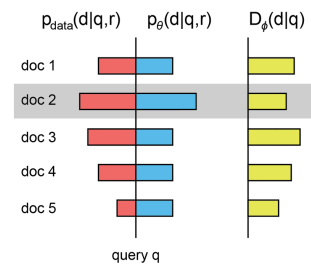

Figure 1: An illustration of GAN for simplest IR tasks where each document is represented just by IDs. For each step, the generative model selects one candidate document for the query and the discriminative model makes a relevance judgment over the query-document pair which is used to guide the generative model for better modeling the user's preference over the candidate documents.

**GANs for Text Generation (35 mins).** Extending from the classic document retrieval tasks, we investigate the potential of leveraging GANs to generate sequences of discrete tokens, where text generation is the most typical scenario [18]. We first reveal the problem of exposure bias of the traditional maximum likelihood (MLE) estimation methods and some previous solutions such as scheduled sampling [2], then we discuss the advantages of GAN techniques for text generation tasks. Many recently proposed techniques will be introduced and compared in this part, including SeqGAN [27], RankGAN [15], LeakGAN [10], MaliGAN [5], GSGAN [13] etc., where SeqGAN is considered as the most representative framework, as illustrated in Figure 2. We will also make an empirical comparison between these models over various evaluation metrics.

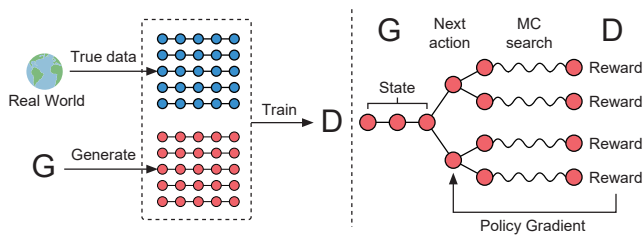

Figure 2: An illustration of SeqGAN for text generation [27]. Compared to one-step generation of discrete token as shown in Figure 1, generating sequence of discrete tokens needs the consideration of subsequent states, which is nationally formulated as a reinforcement learning problem. Here Monte Carlo search is leveraged to make an unbiased estimation of state-action value following the current generation policy.

**GANs for Graph/Network Learning (15 mins).** Graph or network is a complicated discrete data format, based on which there are various information retrieval related tasks like item recommendation, social network analysis, fraud detection, knowledge graphs etc. In this part, we will first briefly introduce network embedding techniques and then introduce the recent work on applying GANs to network embedding and the subsequent graph mining tasks like GraphGAN [23] and KBGAN [4]. The key difference of applying GANs on graph data and sequence data lies in the token (node) sampling strategies as illustrated in Figure 3. The solutions on graph data make use of the graph neighborhood information to higher the sampling efficiency.

**Future Prospective and Summarization (20 mins).** At the end of the tutorial, we raise several unsolved problems and promising future research directions of GANs for IR tasks, e.g., cooperative training (CoT) [17] and generation evaluation methods [29], are discussed. We finally summarize this tutorial by highlighting the key factors and taking home messages of this lecture.

**Q&A Session (15 mins).** We communicate with the audience by question answering, where we could also share some experience of conducting experiments of GANs and reinforcement learning techniques for IR tasks.

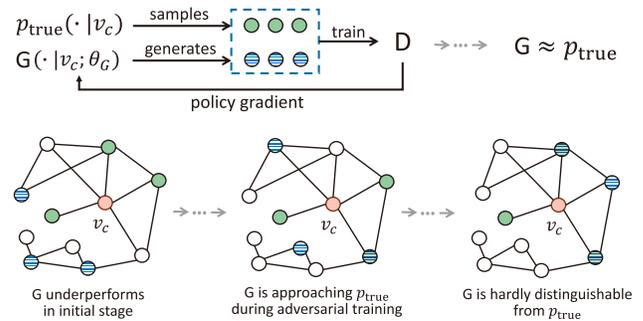

Figure 3: An illustration of GraphGAN for neighbor node generation (selection) [23]. Given a target node $v_c$, for each step the generative model selects another node which is supposed to be linked to $v_c$ and the discriminative model is trained to distinguish the real node pairs and the generated ones.

## 4 SUPPORTING MATERIALS
### 4.1 Online Meterials
A website (http://wnzhang.net/tutorials/sigir2018/) for this tutorial will be made available online right before the lecture is presented. All the relevant materials will be made available on this website, including the talk information, presentation slides, referred papers, speaker information and related open source projects etc.

### 4.2 Offline Materials
We may prepare paper handouts for each audience in the lecture. This depends on the cost of printing in Ann Arbor.

### 4.3 Open Sourced Platforms
Some open sourced platforms will be introduced to the audience to encourage them to run simple code of GANs for IR tasks.
- **IRGAN** [24] (https://github.com/geek-ai/irgan), a code framework of GANs for discrete data generation based on REINFORCE algorithms to support web search, item recommendation and question answering tasks.
- **Texygen** [29] (https://github.com/geek-ai/Texygen), a benchmarking platform, to support research on open-domain text generation models. Texygen has not only implemented a majority of text generation models, but also covered a set of metrics that evaluate the diversity, the quality and the consistency of the generated texts.
- **GraphGAN** [23] (https://github.com/hwwang55/GraphGAN), a repository of the recent solutions for learning node embedding, link prediction and node classification based on GAN techniques.

## 5 RELEVANCE TO THE INFORMATION RETRIEVAL COMMUNITY
Apparently, this tutorial is highly relevant to SIGIR 2018 referred topics and the information retrieval community. As described in Section 3.2, the core studied problem is the deep generative and discriminative models in information retrieval, which is highly relevant to *Artificial Intelligence, Semantics, and Dialog* track in SIGIR

2018; the discussed application scenarios on which the machine learning models are based including web search, recommender systems, question answering, text composing, social network analysis etc., which are the cared topics in *Search and Ranking* track and *Content Recommendation, Analysis and Classification* track; the overall adversarial learning framework itself is a novelty, which would be insightful to the audience from *Foundations and Future Directions* track.

Past relevant tutorials in top-tier conferences include:

- Hang Li, Zhengdong Lu. Deep Learning for Information Retrieval. SIGIR 2016.
- Tom Kenter, Alexey Borisov, Christophe van Gysel, Mostafa Dehghani, Maarten de Rijke, Bhaskar Mitra. Neural Networks for Information Retrieval (NN4IR). SIGIR 2017.
- Bhaskar Mitra, Nick Craswell. Neural Text Embeddings for Information Retrieval. WSDM 2017.

Although these tutorials cover deep learning techniques and their applications on IR, they focus on discriminative models or single generative models rather than adversarial training mechanisms for IR. To our knowledge, this is the first tutorial focused on GAN techniques for IR.

## 6 SPEAKER INFORMATION AND QUALIFICATION

The tutorial speaker, Dr. Weinan Zhang, is current a tenure-track assistant professor in Shanghai Jiao Tong University. His research interests include machine learning and big data mining, particularly, deep learning and reinforcement learning techniques for real-world data mining scenarios, such as computational advertising, recommender systems, text mining, web search and knowledge graphs. He was selected as one of the 20 rising stars of KDD research community in 2016. His work IRGAN won SIGIR 2017 Best Paper Honorable Mention Award. He and his teammates won the 3rd place in KDD-CUP 2011 for Yahoo! Music recommendation challenge and the final champion in 2013 global RTB advertising bidding algorithm competition. Weinan earned his Ph.D. from University College London in 2016 and Bachelor from ACM Class of Shanghai Jiao Tong University in 2011. He was an intern at Google, Microsoft Research and DERI.

As for presenting such a tutorial on GANs for IR, Weinan Zhang would be qualified because (i) he is one of the major authors of IRGAN [24], which is the first fundamental research work on solve IR problems with a GAN solution; (ii) he leads the research projects of SeqGAN [27], LeakGAN [10] and Texygen [29], which are among the most advanced researches for text generation; (iii) he is also an author of GraphGAN [23], which is the first work on node embedding learning based on graph data.

Besides, Weinan Zhang has led or participated several tutorials:

- Weinan Zhang, Jian Xu. Learning, Prediction and Optimisation in RTB Display Advertising. CIKM 2016.
- Jun Wang, Shuai Yuan, Weinan Zhang. Real-Time Bidding based Display Advertising: Mechanisms and Algorithms. ECIR 2016.

**Acknowledgment.** The work is sponsored by National Natural Science Foundation of China (61632017, 61702327, 61772333), Shanghai Sailing Program (17YF1428200), MSRA Collaborative Research and DIDI Collaborative Research Funds.